\documentclass{article}
\usepackage[dvips]{graphicx}  
\usepackage{color}

\usepackage[varg]{txfonts}
\usepackage{concmath}

\begin{document}

\newcommand{\rum}{\rule{0.5pt}{0pt}}
\newcommand{\rub}{\rule{1pt}{0pt}}
\newcommand{\rim}{\rule{0.3pt}{0pt}}
\newcommand{\numtimes}{\mbox{\raisebox{1.5pt}{${\scriptscriptstyle \rum\times}$}}}
\newcommand{\numtimess}{\mbox{\raisebox{1.0pt}{${\scriptscriptstyle \rum\times}$}}}
\newcommand{\Boldsq}{\vbox{\hrule height 0.7pt
\hbox{\vrule width 0.7pt \phantom{\footnotesize T}%
\vrule width 0.7pt}\hrule height 0.7pt}}
\newcommand{\two}{$\raise.5ex\hbox{$\scriptstyle 1$}\kern-.1em/
\kern-.15em\lower.25ex\hbox{$\scriptstyle 2$}$}

\renewcommand{\refname}{References}
\renewcommand{\tablename}{\small Table}
\renewcommand{\figurename}{\small Fig.}
\renewcommand{\contentsname}{Contents}

\twocolumn[%
\begin{center}
{\Large\bf 
Dynamical 3-Space: neo-Lorentz Relativity\rule{0pt}{13pt}}\par

\bigskip
Reginald T. Cahill \\ 
{\small\it  School of Chemical and Physical  Sciences, Flinders University,
Adelaide 5001, Australia\rule{0pt}{15pt}}\\
\raisebox{+1pt}{\footnotesize E-mail: Reg.Cahill@flinders.edu.au}\par

\bigskip

{\small\parbox{11cm}{%
The major extant relativity theories - Galileo's Relativity (GaR),  Lorentz's  Relativity (LR)  and Einstein's Special Relativity (SR), with the latter much celebrated, while the LR is  essentially ignored. Indeed it is often incorrectly claimed that SR and LR  are experimentally indistinguishable.   Here we show that (i) SR and LR  are experimentally distinguishable, (ii)  that comparison of gas-mode Michelson interferometer experiments with spacecraft earth-flyby Doppler shift data demonstrate that it is  LR that is consistent with the data, while SR is in conflict with the data, (iii) SR is exactly derivable from GaR by means of a mere linear change of  space and time coordinates that mixes the Galilean space and time coordinates. So it is GaR and SR that are equivalent. Hence the well-known SR relativistic effects are  purely coordinate effects, and cannot correspond to the observed relativistic effects.  The connections between these three relativity theories has become apparent following the discovery that space is an  observable  dynamical textured system, and that  space and  time are distinct phenomena, leading to a neo-Lorentz Relativity (nLR).  The observed relativistic effects are dynamical consequences of nLR and 3-space.  In particular a proper derivation of the Dirac equation from nLR is given, which entails  the derivation of the rest mass energy $mc^2$.
\rule[0pt]{0pt}{0pt}}}\medskip
\end{center}]{%

\setcounter{section}{0}
\setcounter{equation}{0}
\setcounter{figure}{0}
\setcounter{table}{0}

\markboth{Cahill R.T.  Dynamical 3-Space: neo-Lorentz Relativity}{\thepage}
\markright{ Cahill R.T. Dynamical 3-Space: neo-Lorentz Relativity}

\tableofcontents

\section{Introduction}  
Physics has failed, from the early days of Galileo and Newton, to consider the existence of space as a structured, detectable and dynamical system, and one that underpins all phenomena, until 2002  when it was discovered that the Michelson-Morley experiment was not null \cite{MMCK,MMC}, and indeed confirmed Lorentz's Relativity Theory\footnote{This report is from the Gravitational Wave Detector Project at Flinders University.}.  Essentially the last 400 years of physics has been  one of much confusion because the key phenomenon of space had been  missed, and indeed Minkowski and Einstein had even denied its existence, claiming instead the actual existence of spacetime, a geometrical amalgam of  the geometrical models of space and time \cite{Mink}. Subsequent to the above discovery the dynamics of space has been determined and subjected to many experimental and observational tests, from laboratory experiments to the discovery of the uniformly expanding universe \cite{Book,Review,Paradigm,EmergentGravity,BlackHoles,Universe,CahillNASA,CahillFractalSpace}.  The discovery of dynamical space changes all of physics.  It is now possible to sort out the confusion over which relativity principle is confirmed by experiment, and for the 1st time we  get a clear picture of the nature of reality.

A ``Relativity Principle" (RP) specifies how observations by different observers are related.  In doing so the RP reflects fundamental aspects of realty, and any proposed RP is subject to ongoing experimental challenge.  

There have been  three major relativity theories: Galileo's Relativity (GaR),  Lorentz's  Relativity (LR)  and Einstein's Special Relativity (SR), with the later much celebrated\footnote{The 100th year of SR was celebrated in the 2005 UN Year for Physics.}, while the LR is  essentially ignored. Yet it is often claimed that they are experimentally indistinguishable.   Here we show that (i) they are experimentally distinguishable, (ii)  that comparison of gas-mode Michelson interferometer experiments with spacecraft earth-flyby Doppler shift data \cite{CahillNASA,And2008} demonstrate that it is  LR that is consistent with the data, while SR is in conflict with the same data, (iii) SR is exactly derivable from Galilean Relativity by means of change of  space and time coordinates, so that the well-known SR relativistic effects are purely coordinate effects, and cannot correspond to the observed dynamical relativistic effects.  The connections between these three relativity theories has become apparent following the discovery that space is a dynamical and observable system, and that  space and  time are distinct phenomena. 

We give a non-historical presentation, because historical presentations were always confused by the lack of realisation that a dynamical space existed, although serious consideration was given to Lorentz Relativity 
\cite{BrownContraction,BrownBook,BrownPooley}.

But 1st a warning: a common error when discussing the physics of space and time is to confuse space and time coordinates with the actual phenomenon of space and time, and also to confuse space intervals, as measured by a ruler or round trip light speed measurements, and time measured by an actual clock, with actual intrinsic measures of space and time phenomena: coordinates are arbitrary, whereas the intrinsic measures are set by the dynamics of space.

\section{Quantum Foam Dynamical Space\label{sect:QFoam}}
 \begin{figure}[t]
\vspace{0mm}
\hspace{2mm}\includegraphics[scale=0.32]{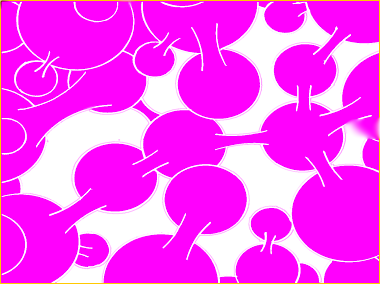}\hspace{2mm}\includegraphics[scale=0.165]{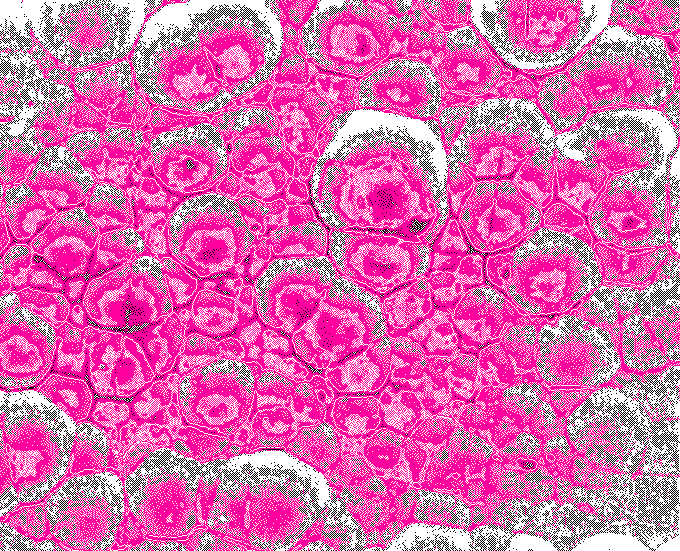}
	\caption{\small Left: Representation of space at a microscopic scale as a quantum foam (Quantum Homotopy Field Theory) - with networks forming connected $S^n$'s shown embedded, at a coarse grain level, in an an emergent $E^3$. The quantum foam is fractal, so this connectivity is repeated at all scales \cite{Book}. 
	Right: Representation of the fractal 3-space velocity field ${\bf v}({\bf r},t)$ as determined by experiment \cite{CahillFractalSpace}.  This detected space structure is passing the earth at $\sim$500km/s, with the velocity slightly different within each  cell. This data implies that dynamical space has a fractal texture.}
\vspace{-4mm}\label{fig:QFoam}\end{figure}

Our understanding of reality evolves through ongoing iterations of theory and experiments/observations.  For the discussions herein we, as always, need an explicit  paradigm, otherwise the words ``space" and ``time" lack meaning. To that end we note that a deeper Quantum Homotopy Field Theory model of space arises from a  stochastic non-quantum information-theoretic account \cite{Book}: this involves an emergent  wave-functional  $\Psi[..,\pi_{\alpha\beta},$ $..,t]$ where the configuration space is that of homotopic mappings between closed compact networks:  $\pi_{\alpha\beta}:S^\alpha\rightarrow S^\beta$, see Fig\ref{fig:QFoam} - these are the Nambu-Goldstone modes of the information-theoretic system. $\Psi[t]$  evolves in time according to a stochastic  functional Schr\"{o}dinger equation, and generates a quantum foam structure.   This wave-functional has an approximate embedability in $E^3$, permitting the use of $E^3$ coordinates ${\bf r}=\{x,y,z\}$. This $E^3$ is an emergent coarse-grained property of the quantum foam, and not a separate entity.  So we are not dealing with a dual system. The model also has an absolute/universal  time index $t$. This  cosmic  time is observable, but which however is not, in general, given by ``clock time".  Quantum matter is described by wave-functional components having topological 
properties - essentially Skyrmionic solitons.  A classical description of this quantum foam is given by a velocity field ${\bf v}({\bf r},t)$,  with the dynamics determined by phenomenological arguments, see Sect.\ref{sect:dynamics}, where the velocity is defined relative to a network of observers, using an $E^3$ to label space points. Experiment and observations have revealed that the space  dynamics is at least  a 3-parameter  system: $G$ - Newton's gravitational constant,  $\alpha$ - the fine structure constant, which determines a space self-interaction effect,  and $\delta$ - having the dimensions of length, describing another self-interaction effect, and which appears to be a very small Planck-like length \cite{BlackHoles}. The fractal structure of the textured space at the macroscopic level has been detected by experiments \cite{CahillFractalSpace}, see Fig.\ref{fig:QFoam}.

\section{Galilean Relativity}
We give here a modern statement of Galilean Relativity from the point of view of the paradigm in Sect.\ref{sect:QFoam}. The assumptions in GaR are (i) space exists, but is not observable and not dynamical, and is modelled as a Euclidean 3-space ($E^3$),  which entails the notion that space is without structure, (ii) observers measure space and time intervals using rods and clocks, whose respective lengths and time intervals are not affected by their motion through space, (iii)  velocities are  measured relative to observers, where different observers, $O$ and $O^\prime$, relate  their space and time coordinates by
\begin{equation}
t^\prime=t,\mbox{\ \ \  }
x^\prime=x-Vt,  \mbox{\ \ \  }y^\prime=y, \mbox{\ \ \  }z^\prime=z.
\label{eqn:Gcoords}\end{equation}
where $V$ is the relative speed of the observers (in their common $x$-direction, for simplicity). The speed $w$ of an object or waveform (in the $x$ direction) according to each observer, is related by
\begin{equation}
w^\prime=w-V
\label{eqn:GSpeedTransf}\end{equation}
Eqns (\ref{eqn:Gcoords}) and (\ref{eqn:GSpeedTransf}) form the Galilean Relativity Transformation, and the underlying assumptions define Galilean Relativity (GaR). Newton based his dynamics on Galilean Relativity, in particular his theory of gravity, to which General Relativity reduces in the limit of low mass densities and low speeds.

\section{Lorentz and Neo-Lorentz Relativity}
When Maxwell formulated his unification of electric and magnetic fields\footnote{The now standard formalism was actually done by Heaviside.} the speed of EM waves came out to be the constant $c=1/\sqrt{\epsilon_0\mu_0}$ for any observer, and so independent of the motion of the observers wrt one another or to space. This overtly contradicted GaR, in (\ref{eqn:GSpeedTransf}).  Hertz  in 1890 \cite{Hertz} pointed out the obvious fix-up, namely that Maxwell had mistakenly not used the then-known Euler constituent derivative $\partial/\partial t +{\bf v}\cdot\nabla$, in place of $\partial/\partial t$, where $\bf v$ is the velocity of some structure to space relative to an observer, in which case Maxwell's equations would only be valid in the  local rest frame defined by this structure.  In that era a dual model was then considered, namely with a Euclidean space $E^3$ and an extended all-filling aether substance, so that the velocity $\bf v$ was the velocity of the aether relative to an observer.
To be explicit let us consider  the case of  electromagnetic  waves, as described by the vector potential ${\bf A}({\bf r},t)$ satisfying the wave equation (in absence of charges and currents), but using the Euler constituent derivative, as suggested by Hertz: 
\begin{equation}
\left(\frac{\partial}{\partial t} +{\bf v}({\bf r},t)\!\cdot\!\nabla \right)^2\!{\!\bf A}({\bf r},t)=c^2\nabla^2{\bf A}({\bf r},t).
\label{eqn:Aeqn}\end{equation}
Here $\nabla=\{\frac{\partial}{\partial x},\frac{\partial}{\partial y},\frac{\partial}{\partial z}\}$. 
In Lorentz Relativity there is a static aether in addition to an actual Euclidean space, so  ${\bf v}$ is independent of ${\bf r}$ and $t$; whereas in neo-Lorentz Relativity ${\bf v}({\bf r},t)$
describes a dynamical space, with ${\bf r}$ and $t$ describing a cosmic embedding space and a cosmic time. 
 We find plane-wave solutions only  for the case where the space flow velocity, relative to an observer,  is locally time and space independent, {\it viz} uniform,
$$
{\bf A}({\bf r},t)={\bf A}_0\sin({\bf k}\cdot{\bf r}-\omega t)
$$
with $\omega({\bf k},{\bf v})=c|\vec{{\bf k}}| +{\bf v}\cdot{\bf k}$.
The EM wave group velocity is then
$$
{\bf v}_g=\vec{\nabla}_k\omega({\bf k},{\bf v})=c\hat{\bf k}+{\bf v}
$$
and we see that the wave has velocity ${\bf v}_g$ relative to the observer, with the space flowing at velocity  ${\bf v}$ also relative  to the observer, and so the EM speed  is $c$ in direction $\hat{\bf k}$ relative to the aether (LR) or space (nLR).   In searching for experimental evidence for the existence of this aether, or more generally a Preferred Frame of Reference (PFR), Michelson conceived of his interferometer, see Appendix \ref{append:MM}.  Unknown to Michelson was that his design had an intrinsic fatal flaw: if operated in vacuum mode it was incapable of detecting the PFR effect, while with air present, as operated by Michelson and Morley in 1887, it was extremely insensitive  \cite{MMCK,MMC}.  The problem was that Michelson had used Newtonian physics, {\it viz} GaR, in calibrating the interferometer, Sect. \ref{subsect:GaR}.  Michelson and Morley detected fringe shifts, but they were smaller than expected, and were interpreted as a null effect:  there was no aether or PFR effect. However Lorentz \cite{Lorentz,Lorentz1892} and Fitzgerald \cite{Fitzgerald} offered an alternative explanation: physical objects, such as the arms supporting the interferometer optical elements, undergo a contraction in the direction of movement through  the aether, or more generally  relative to the PFR: the length becoming $L= L_0 \sqrt{1-v^2_R/c^2}$, where $L_0$ is the physical length when at rest wrt the PFR, and $v_R $ is the speed relative to the PFR.  It must be noted that this is not the Lorentz contraction effect predicted by SR, as discussed later, as that involves $L= L_0 \sqrt{1-v^2_O/c^2}$, where $v_O$ is the speed of the arm or {\it any} space interval relative to the observer.  The difference between these two predictions is stark,   and has been observed experimentally, and the SR prediction is proven wrong, see Sects. \ref{subsect:SRCalibA} and  \ref{subsect:SRCalibB}.

Next consider two observers, $O$ and $O^\prime$, in relative motion. Then the actual intrinsic or physical time and space coordinates of each are, in both LR and nLR, related by the Galilean transformation, and here we consider only a uniform ${\bf v}$:  these coordinates are not the directly measured  distances/time intervals - they require corrections to give the intrinsic values.
We have taken the simplest case where $V$ is the intrinsic relative speed of the two observers   in their common $x$ directions.   Then from (\ref{eqn:Gcoords}) the derivatives  are related by
$$
\frac{\partial}{\partial t}=\frac{\partial}{\partial t^\prime}-V\frac{\partial}{\partial x'}, 
\frac{\partial}{\partial x}=\frac{\partial}{\partial x^\prime}, \frac{\partial}{\partial y}=\frac{\partial}{\partial y^\prime}, \frac{\partial}{\partial z}=\frac{\partial}{\partial z^\prime}.
$$
In the general case space rotations may be made.Then (\ref{eqn:Aeqn})   becomes for the 2nd observer, with $v^\prime=v-V$,
\begin{equation}
\left(\frac{\partial}{\partial t^\prime} +{\bf v}^\prime\!\cdot\!\nabla^\prime \right)^2\!\!{\bf A}^\prime({\bf r}^\prime,t^\prime)=c^2\nabla^{\prime 2}{\bf A}^\prime({\bf r}^\prime,t^\prime).
\label{eqn:EM}\end{equation}
with  ${\bf A^\prime}({\bf r^\prime},t^\prime)={\bf A}({\bf r},t)$.
If the flow velocity  ${\bf v}({\bf r},t)$ is not uniform then we obtain refraction effects for the EM waves, capable of producing   gravitational lensing. Only for an observer at rest in a time independent and uniform  aether (LR) or space (nLR)  does $v^\prime$  disappear from (\ref{eqn:EM}).

\section{Special Relativity from Galilean \newline Relativity}
The above  uses physically intrinsic  choices for the time and space coordinates, which are experimentally accessible.   However we could   choose to use a new class of time and space coordinates, indicated by upper-case symbols $T,X$, $Y$, $Z$, that mixes the above time and space coordinates. We begin by showing  that Special  Relativity (SR), with its putative spacetime as the foundation of reality, is nothing more than Galilean Relativity (GaR) written in terms of these mixed space and time coordinates.  The failure to discover this, until 2008 \cite{CahillMink} reveals one of the most fundamental blunders in physics.  One  class of such mixed coordinates for $O$ is\footnote{It is important to use different notation for the GaR coordinates and the SR coordinates: often the same notation is used, illustrating the confusion in this subject.}
\begin{eqnarray}
T&=&\gamma(v)\left((1-\frac{v^2}{c^2})t+\frac{v x}{c^2}\right),\nonumber \\ X&=&\gamma(v)x, \mbox{\   }Y=y, \mbox{\   }Z=z
\label{eqn:mixedST}\end{eqnarray}
where $v$ is the uniform speed of space (in the $x$ direction), and where $\gamma(v)=1/\sqrt{1-v^2/c^2}$. Note that this is not a Lorentz transformation. 
If an object has speed $w$, $x=wt$, wrt to $O$,  then it has speed $W$, $X=WT$, using the mixed coordinates, wrt $O$ 
\begin{equation}
W=\frac{w}{1-\frac{v^2}{c^2} +\frac{v}{c^2}w}
\label{eqn:Ww}\end{equation}
Similarly for $O'$ using $v'$, $w'$ and $W'$.  In particular (\ref{eqn:Ww}) gives for the relative speed of $O'$ wrt $O$ in the mixed coordinates 
\begin{equation}
\overline{V}=\frac{V}{1-\frac{v^2}{c^2} +\frac{v}{c^2}V}
\label{eqn:VbarV}\end{equation}
Using the above we may now express the Galilean speed transformation (\ref{eqn:GSpeedTransf}) in terms of $W',W$ and $\overline{V}$ for the mixed coordinates,   giving
\begin{equation}
W'=\frac{W-\overline{V}}{1-W\overline{V}/c^2}
\label{eqn:SRVelTran}\end{equation}
which is the usual SR transformation for speeds, but here derived exactly from the Galilean transformation. Note that
 $c$ enters here purely because of the definitions in  (\ref{eqn:mixedST}), which is designed to ensure that wrt the mixed space-time coordinates the speed of light is invariant: $c$. To see this 
note that from (\ref{eqn:mixedST}) the transformations for the derivatives are  found to be
\begin{eqnarray}
\frac{\partial}{\partial t}&=&\gamma(v)\left(1-\frac{v^2}{c^2}\right)\frac{\partial}{\partial T}, \nonumber \\
\frac{\partial}{\partial x}&=&\gamma(v)\left(\frac{v}{c^2}\frac{\partial}{\partial T}+\frac{\partial}{\partial X}\right), \nonumber \\
  \frac{\partial}{\partial y}&=&\frac{\partial}{\partial Y},  \mbox{\ \ }\frac{\partial}{\partial z}=\frac{\partial}{\partial Z}.
\end{eqnarray}
 $\overline{\nabla}=\{\frac{\partial}{\partial X},\frac{\partial}{\partial Y},\frac{\partial}{\partial Z}\}$. Then we have from (\ref{eqn:Aeqn}), for uniform $v$,
$$
\left(\frac{\partial}{\partial T}  \right)^2\!\!\overline{{\bf A}}({\bf R},T)=c^2\overline{\nabla}^2\overline{{\bf A}}({\bf R},T).
$$
with ${\bf R}=\{X,Y,Z\}$ and $\overline{\bf A}({\bf R},T)={\bf A}({\bf r},t)$. The speed of EM waves is now $c$ for all observers. This is a remarkable result. In the new class of coordinates the dynamical equation no longer contains the space velocity  $\bf v$ - it has been mapped out of the dynamics. The EM dynamics  is now invariant under Lorentz transformations.
\begin{eqnarray}T^\prime&=&\gamma(\overline{V})\left(T-\frac{\overline{V}X}{c^2}\right),  \nonumber \\
X^\prime&=&\gamma(\overline{V})(X-\overline{V}T), \mbox{\ }Y^\prime=Y,  \mbox{\ \ \  }Z^\prime=Z,
\label{eqn:LT}\end{eqnarray}
and we note that for two events with coordinate differences \newline $\{dT, dX\}$ or  $\{dT^\prime, dX^\prime\}$ 
\begin{equation}
dI^2\equiv c^2dT^{\prime 2}-dX^{\prime 2}=c^2dT^2-dX^2
\label{eqn:invariantinterval}\end{equation} 
defines the   invariant interval for different observers.
 There is now no reference to the underlying flowing space: for an observer using this class of space and time coordinates the speed of EM waves  relative to the observer is  always $c$ and so invariant - there will be no EM speed  anisotropy.  We could also introduce, following Minkowski,   ``spacetime"  light cones along which $d\tau^2=dT^2-d{\bf R}^2/c^2=0$.  Note that $d\tau^2$ is invariant under the Lorentz transformation (\ref{eqn:LT}). Then pairs of spacetime events could be classified into either time-like, $d\tau^2>0$, or space-like, $d\tau^2\leq0$, with the time ordering of spacelike events not being uniquely defined.  However this outcome is merely an artifact of the mixed space-time coordinates: $dT$ is not the actual time interval.

\begin{figure}
\vspace{-10mm}
\setlength{\unitlength}{0.75mm}
\hspace{15mm}
\begin{picture}(0,30)
\thicklines

\definecolor{hellgrau}{gray}{.8}
\definecolor{dunkelblau}{rgb}{0, 0, .9}
\definecolor{roetlich}{rgb}{1, .7, .7}
\definecolor{dunkelmagenta}{rgb}{.9, 0, .0}
\definecolor{green}{rgb}{0, 1,0.4}
\definecolor{black}{rgb}{0, 0, 0}
\definecolor{dunkelmagenta}{rgb}{.9, 0, .0}

 \put(36,4){ $O$}
  \put(46,4){ $W$}
    \put(60,4){ $O^\prime$}
\color{dunkelblau}

\put(10,0){\line(1,0){30}}
\put(10,1){\line(1,0){30}}
\put(40,0.5){\vector(1,0){10}}
\put(10,0){\line(0,1){1}}
\put(40,0){\line(0,1){3}}

    \put(64,0){\line(0,1){3}}

\end{picture}
\caption{\small  Here is derivation of SR length contraction from Galilean Relativity using coordinates introduced in (\ref{eqn:mixedST}). Consider two events: (1) RH end of rod travelling with observer $O$,  with speed $W$ wrt observer $O^\prime$,  passes $O^\prime$,    and (2)  when LH end passes $O^\prime$.   Then $dX^\prime=0$, and  $L^\prime= WdT^\prime$ defines $L^\prime$. For $O$   $dX= L$ and $L=WdT$. Then 
(\ref{eqn:invariantinterval}) gives  $L^\prime=\sqrt{1-W^2/c^2}L$, with $W$ the speed of the rod wrt $O^\prime$. However this is purely a coordinate effect, and has no physical significance. Experiment shows that it is the speed of the rod wrt space, $v_R$, that actually determines the length contraction.}
\label{fig:length}\end{figure}
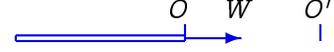

\begin{table*}[t]
\caption{\large Space and  Relativity Theories}  
\centering
\begin{tabular}{|l|l|l|l|l|l|}
\hline
Theory & Embedding Space? &  Dynamical Space?  & Experiments &  Comments \\  \hline\hline
\small{Galilean} & \small{No. Space is a Real }  & \small{No. Later Newton }& \small{No absolute motion effects} & \small{Absolute Euclidean space} \\  
  \small{(GaR)       1632     }                   & \small{Euclidean space.} & \small{suspects aether flow. }   & \small{Disagrees with later expts.}  &\small{and absolute time.}\\  \hline

\small{Lorentz}  & \small{No. Real  Euclidean } &  \small{No (static aether).  } & \small{Absolute motion effects. }& \small{Dual system: absolute }\\  
\small{(LR)   1892 }                       & \small{Space. }  &  \small{No dynamics. }  & \small{1st detected 1887.}&  \small{embedding  space \& aether. }\\  \hline

\small{Einstein}  & \small{No. Geometric space  } & \small{No.  Dynamical spacetime }  & \small{No real dynamical effects -} &\small{Equivalent to GaR. }  \\  
       \small{(SR)  1905     }               &   \small{and time amalgamated into }&  \small{in  GR - but requires dark }  &\small{only  coordinate effects. }&  \small{Absolute spacetime is} \\    &\small{ real spacetime.} & \small{matter and dark energy.}& \small{Disagrees with expt. } & mathematical artefact.\\  \hline

\small{neo-Lorentz } &\small{Yes.  Emergent property}  &  \small{Yes. From quantum} & \small{Dynamical space 1st detected}& \small{Inequivalent to SR  \& GaR.  }\\  
   \small{(nLR)       2005\cite{Book}  }                & \small{Euclidean or compact. } &\small{foam.  Causes gravity} &\small{1887. Relativistic effects}  &  \small{Generalisation of LR. }\\  
   &\small{Provides coordinate labels } &\small{and light bending.}&  \small{from absolute motion.}   & \small{Absolute dynamical-space }\\  
     &\small{for dynamical space. }  &  \small{Known dynamics: $G, \alpha, \delta$.}   & \small{$\alpha \approx 1/137$ from grav. expts. } &   \small{and absolute time. } \\  \hline
\end{tabular}
\label{tab:summary}\end{table*}
}

Confusing a space and time coordinate system with actual  space and time phenomena has confounded physics for more than 100 years, with this illustrated above by the recently discovered  exact relationship between Galilean Relativity and Einstein Relativity.
In mainstream physics it is claimed that Special Relativity reduces to Galilean Relativity only in the limit of speeds small compared to $c$.
But the various so-called ``relativistic effects"  ascribed to Special Relativity are nothing more than coordinate effects - they are not real.  It was Lorentz who first gave a possible {\it dynamical} account of relativistic effects, namely that they are caused by absolute motion of objects relative to  the aether (LR) or, now, dynamical space (nLR), which according to the evidence discussed above, is absolute motion relative to a dynamical and structured quantum foam substratum: space.  In Lorentz Relativity relativistic effects are genuine dynamical effects and must be derived from some dynamical theory. This has yet to be done, and for the length contraction effect would involve the quantum theory of matter.  

Finally we note in Fig.\ref{fig:length} that the so-called length contraction effect in SR is exactly derivable from GaR - and so it is purely a coordinate effect, and so has no physical meaning.

\section{Detecting Lorentz Relativistic Effects}
We now show how only  Lorentz Relativity gives a valid account  of the experimental results dealing with light  speed anisotropy.  To that end we consider the differing predictions made by the relativity theories for the length contraction effect,  and    we use data from Michelson interferometer experiments, which being a 2nd order in $v/c$ detector  requires length contraction effects to be included, when relevant.   These contradictory predictions are compared with detailed data from the NASAspacecraft earth-fly Doppler shifts, which in LR and nLR do not involve any length contraction, as no objects/supporting arms are involved. The flyby Doppler shifts have been also confirmed by laboratory 1st order $v/c$ experiments by  DeWitte and Cahill, see \cite{CahillFractalSpace}, and so not requiring 2nd order $v/c$ length contraction effects to be considered.
 So we have a critical and decisive test of the relativity theories.  In all cases we parametrise the calibration theory for the Michelson interferometer  travel time difference between the two arms according to
 \begin{equation}
 \Delta t=k^2\frac{L_0v_p^2}{c^3}\cos(2\theta)
 \label{Mcalb}\end{equation}
 where $k^2$ is the theory-dependent calibration constant. Here $L_0$ is the at-rest arm length, $v_P$ is the relevant velocity  projected onto the plane of the interferometer, and $\theta$ is the angle between that projected velocity and one of the arms, see Fig.\ref{fig:Minterferometer}.

\subsection{Lorentz and neo-Lorentz Relativity \newline Interferometer Calibration}

In both LR and nLR the length contraction effect is a real dynamical effect caused by the absolute motion of an actual object wrt aether (LR) or dynamical space (nLR).  In  Appendix \ref{append:MM} we give a simple analysis that yields the calibration constant  $k^2=(n^2-1)$, when $n\approx 1$ is the refractive index of the gas present: for air $n=1.00029$ at STP, giving $k^2=0.00058$. Some data from the Michelson-Morley and Miller experiments are shown in Fig.\ref{fig:MandMM}, showing, together with other data,  that this value of $k^2$ gives excellent agreement with the Doppler shift data, and  different 1st order in $v/c$ experiments  
\cite{CahillFractalSpace}.  The gas-mode interferometer experiments  and spacecraft Doppler shift data give $v\approx 500$km/s. Note that high-accuracy vacuum-mode Michelson interferometers will give a null result ($n=1$), as has been repeatedly observed.

\subsection{Galilean Relativity  Interferometer \newline Calibration\label{subsect:GaR}}
In Galilean Relativity there is no length contraction effect, and repeating the analysis in  Appendix \ref{append:MM}, without that effect,  we obtain $k^2=n^3$ ($\approx 1$ for air). This is the calibration constant used by Michelson-Morley in 1887.  Using this to analyse their data  they found that $v_P \leq 10$km/s.  This is in stark conflict with the speed of $v\approx 500$km/s from spacecraft earth-flyby Doppler shift and 1st order in $v/c$ experiments.  So Galilean Relativity is ruled out.

\subsection{A: Einstein Relativity Interferometer \newline Calibration \label{subsect:SRCalibA}}
There are two routes to $k^2$ from Einstein Relativity, depending on which choice of space and time variables is used.  Here we use the Galilean space and time coordinates, as we have shown that they are the physical coordinates that underly SR, in which case $k^2=n^3$, giving $v_P \leq 10$km/s as in Sect. \ref{subsect:GaR}, and so again is in stark disagreement with experimental data.

\subsection{B: Einstein Relativity Interferometer \newline Calibration \label{subsect:SRCalibB}}
In this approach we use the mixed space and time coordinates conventionally used in SR calculations. Then the speed of light is $c/n$ - invariant wrt to these coordinates, but there is no length contraction effect, because the arms are at rest wrt the observer.  Then again we find that  $k^2=n^3\approx 1$,
and in disagreement with the experimental data.

\section{Dynamical 3-Space and \newline Neo-Lorentzian  Relativity\label{sect:dynamics}}
We briefly outline the dynamical modelling of 3-space. It involves the space velocity field ${\bf v}({\bf r},t)$, defined relative to an observer's frame of reference, and using intrinsic space and time coordinates,  
\begin{equation}
 \nabla\!\!\cdot\!\!\left(\!\!\frac{\partial\mathbf{v}}{\partial t}\!\!+\!\!\left(\mathbf{v}\!\cdot\!\nabla\!\right)\mathbf{v}\!\!\right)\!+\!
 \frac{\alpha}{8}\!\!\left(\!\left(\mathrm{tr}D\right)^{2}\!-\!\mathrm{tr}\!\left(D\right)^{2}\right)\!\! +..=-4\pi G\rho
\label{eqn:3spaceequation}
\end{equation}
$\nabla \times \mathbf{v}=\mathbf{0}$ and  $D_{ij}=\partial v_i/\partial x_j$. The velocity field ${\bf v}$  describes classically the time evolution of the substratum quantum foam. The bore hole $g$ anomaly data has revealed $\alpha = 1/137$,  the fine structure constant.
The matter acceleration is found by determining the trajectory of a quantum matter wavepacket \cite{CahillSE}. More generally we can  vary the path ${\bf r}_o(t)$ to maximise the proper travel time $\tau$:
\begin{equation}
\tau=\int dt \sqrt{1-\frac{{\bf v}^2_R({\bf r}_0(t),t)}{c^2}}
\label{eqn:propertime}\end{equation} 
where ${\bf v}_R({\bf r}_o(t),t) ={\bf v}_o(t) - {\bf v}({\bf r}_o(t),t),$ is the velocity of the wave packet, at position ${\bf r}_0(t)$,  wrt the local space - a neo-Lorentzian Relativity effect.
 This ensures that quantum waves propagating along neighbouring paths are in phase, and so interfere constructively.  This proper time expression, which entails that the maximum speed wrt space is $c$,  is  derived in Appendix \ref{append:propertime}. There it follows from the dynamical length contraction, which is yet to be derived.
This maximisation gives the quantum matter geodesic equation for ${\bf r}_0(t)$ \cite{Book}.
\begin{equation}
{\bf g}\!=\!\displaystyle{\frac{\partial {\bf v}}{\partial t}}\!+\!({\bf v}\!\cdot\!{\bf \nabla}){\bf
v}\!+\!({\bf \nabla}\!\times\!{\bf v})\!\times\!{\bf v}_R-\frac{{\bf
v}_R}{1\!-\!\displaystyle{\frac{{\bf v}_R^2}{c^2}}}
\frac{1}{2}\frac{d}{dt}\left(\frac{{\bf v}_R^2}{c^2}\right)+...
\label{eqn:acceleration}\end{equation}  
with  ${\bf g}\equiv d{\bf v}_o/dt=d^2{\bf r}_o/dt^2$.  
The 1st term in $\bf g$ is  the Euler space acceleration   $\bf a$, the 2nd term explains the Lense-Thirring effect, when the vorticity is non-zero,  and the last term   explains the precession of orbits.
The above reveals gravity to be an emergent phenomenon where quantum matter waves are refracted by the time dependent and inhomogeneous 3-space velocity field. The $\alpha$-term in (\ref{eqn:3spaceequation}) explains the so-called ``dark matter" effects:  if $\alpha\rightarrow 0$ and $v_R/c\rightarrow 0$ we derive Newtonian gravity, for then $\nabla\cdot{\bf g}=-4\pi G \rho$ \cite{Book}. Note that the relativistic term in (\ref{eqn:acceleration}) arises from the quantum matter dynamics - not from the space dynamics, which does not involve $c$. Dynamical 3-space theory has been tested in laboratory light speed anisotropy measurements,  $G$ measurement anomalies, bore hole $g$ anomaly, spacecraft earth-flyby Doppler shifts, spiral galaxy flat rotation curves, galactic black holes,  galactic black hole - star motion correlations, galactic light lensing, universe expansion dynamics, universe cosmic textures and  filaments, and gravitational wave detection experiments.

The dynamics of space necessarily involves a Frame of Reference (FoR), by which an observer codifies the location ${\bf r}$  and time $t$  of an event. There is a natural and fundamental choice for these, although in principle any non-degenerate choice is possible.  The natural choice is an absolute FoR defined  by (i) spatial separations $|{\bf r}_1-{\bf r}_2|$ correspond to the intrinsic distance between elements of space, and (ii) time intervals $t_2-t_1$ are absolute cosmic time intervals.  Such a FoR is used in (\ref{eqn:3spaceequation}).  Absolute cosmic time is now observable - it is the time steps defined by a clock that is at rest wrt the local space, so that the clock suffers no dynamical slowing effect.  This merely requires a light speed anisotropy detector:  the measured space speed permits the compensation of a clock moving wrt space. Of course these coordinates are subject to the usual arbitrariness in choices of units, such as {\it meter} and {\it second}.  Globally the embedding space may be an $E^3$, as in (\ref{eqn:3spaceequation}), or an $S^3$, requiring a generalisation of (\ref{eqn:3spaceequation}), etc.   At a deeper level the dynamics of space and quantum matter appears to emerge in a self-organising process from the information-theoretic  {\it Process Physics} \cite{Book}, with the classical physics dynamics in  (\ref{eqn:3spaceequation}) arising from a derivative expansion,  of the underlying Quantum Homotopic Field Theory. This information-theoretic model is a stochastic pattern formation and recognition system, with the stochasticity representing a limitation to self-referencing (SRN - Self Referential Noise), and from which space emerges, leading to the phenomenon of` ``locality" and ``local dynamics", but is intrinsically non-local and highly interconnected.

\section{Deriving Generalised Dirac  and \newline Schr\"{o}dinger Equations}
Because Special Relativity is simply Galilean Relativity using mixed space-time change of coordinates, the well-known SR relativistic effects  cannot be actual physical phenomena, as discussed above.  This implies that the Dirac equation, which historically was constructed starting with SR, must in fact have a different origin. Here we derive the Dirac equation from neo-Lorentz Relativity, and which gives a generalised form that takes account of the existence of the dynamical space. The Schr\"{o}dinger and Dirac wave equations model quantum matter as a wave phenomena, with the particulate aspect of matter, i.e. localised matter,  accounted for by localised wave packets.   The velocity of a  wave packet  is given by the group velocity ${\bf v}_0=\nabla_{\bf k} \omega({\bf k})|_{\bf K}$, wrt an inertial observer, where $\omega({\bf k})$ is the dispersion relation, and ${\bf K}$ is the dominant wave vector that characterises the wave packet.   
\begin{equation}
\psi({\bf r},t)=\int d^3k f({\bf k})e^{-i\omega({\bf k})t+i{\bf k}\cdot{\bf r}},
\label{eqn:wp}\end{equation}
where $ f({\bf k})$ is peaked about ${\bf k}={\bf K}$.

Hence the task is to construct a free-fall wave function equation that  gives the Galilean Relativity energy relation $E=\hbar \omega({\bf K})$ $ =m v_O^2/2+E_0$ for a free wave packet, at low speeds, and which demands $v_R < c$ in the limit of high speeds, where $v_R$ is the speed relative to space,  and so in accord with  (\ref{eqn:propertime}). Here $E_0$ is the  energy of a wave packet at rest wrt to space, which emerges below from neo-Lorentz relativity.    We begin with an ansatz for the time evolution of $\psi({\bf r},t)$:
\begin{equation}
i\hbar\frac{\partial  \psi}{\partial t}=(b^2+a^2(-i\hbar\nabla)^2))^n\psi-i\hbar\left({\bf
v}.\nabla+\frac{1}{2}\nabla.{\bf v}\right)\psi
\label{eqn:PDE1}\end{equation}
where $a$ and $b$ are real numbers to be determined, along with  the exponent $n$.   The  term ${\bf
v}.\nabla$ arises from introducing the  Euler constituent time derivative, which ensures that it is motion wrt the dynamical space that produces dynamical effects. The last term, $\frac{1}{2}\nabla.{\bf v}$, is uniquely determined as the Euler derivative, by itself, does not lead to a hermitian operator on the RHS, needed to produce a unitary time evolution.
For a uniform space flow ${\bf v}$ wrt an observer, moving uniformly through space,   and for  a plane wave solution $\psi=e^{-i\omega({\bf k}) t +i{\bf k}\cdot{\bf r}}$, we obtain
\begin{equation}
\hbar\omega({\bf k})=(b^2+a^2\hbar^2k^2)^n+\hbar{\bf k}\cdot{\bf v}
\label{eqn:PDE2}\end{equation}
The  velocity ${\bf v}_0$ is then
\begin{equation}
{\bf v}_o=\nabla_{\bf k} \omega({\bf k})=2na^2\hbar {\bf k}(b^2+a^2\hbar^2k^2)^{n-1}+{\bf v}
\label{eqn:PDE3}\end{equation}
with ${\bf v}_R={\bf v}_o-{\bf v}$.
For large $k$ \hspace{2mm}$v_R$ has the limiting value $c$ only if $n=1/2$ and  $a=c$.  For small $k$ we then obtain 
\begin{equation}
E=\hbar\omega(k) = b+\frac{\hbar^2k^2c^2}{2b}+\hbar{\bf k}\cdot{\bf v}
\label{eqn:PDE4}\end{equation}
which produces the Galilean Relativity kinetic energy term  \newline $p^2/2m$ \hspace{1mm}  (${\bf p}=\hbar {\bf k}$) only if $b=mc^2$. We thus obtain the rest-mass energy $E_0=mc^2$, when the wave packet is at rest wrt space, and not wrt the observer.  Eqn.(\ref{eqn:PDE1}) is then a generalised spin-$0$ Schr\"{o}dinger  equation
\begin{equation}
i\hbar\frac{\partial  \psi}{\partial t}=mc^2\left(1+\frac{(-i\hbar\nabla)^2)}{m^2c^2}\right)^{1/2}\psi-i\hbar\left({\bf
v}.\nabla+\frac{1}{2}\nabla.{\bf v}\right)\psi
\label{eqn:PDE5}\end{equation}
which for  long wavelengths 
becomes  
\begin{equation}
i\hbar\frac{\partial  \psi({\bf r},t)}{\partial t}=H(t)\psi({\bf r},t),
\label{eqn:SEa}\end{equation}
with the free-fall hamiltonian 
\begin{equation}
H(t)=-\frac{\hbar^2}{2m}\nabla^2-i\hbar\left({\bf
v}.\nabla+\frac{1}{2}\nabla.{\bf v}\right)+mc^2
\label{eqn:SEb}\end{equation}
generalised to now account for the interaction of the wave function with the dynamical space which, as shown above, gives the emergent phenomenon of gravity. The rest mass energy is also present.
Eqn. (\ref{eqn:PDE5}) for plane waves and uniform space flow gives
\begin{equation}
E_p=\hbar \omega - {\bf p}_R\cdot{\bf v}=mc^2\left(1+\frac{v_R^2}{c^2}\right)^{1/2}
\label{eqn:energy}\end{equation}
where ${\bf p}_R=\hbar {\bf k}=m{\bf v}_R$. The RHS is independent of the observer's speed through space, and so $E_p$ is an intrinsic measure of the energy of the particle.   This implies that the non-relativistic limit  kinetic energy of a particle should be  $K=\frac{1}{2}m v^2_R$, which is observer independent.

To include spin we follow the Dirac idea and generalise  (\ref{eqn:PDE1}) by introducing matrices  $B$ and $A_i$
\begin{equation}
i\hbar\frac{\partial  \psi}{\partial t}=(B-i\hbar\vec{A}\cdot\nabla)^{2n}\psi-i\hbar\left({\bf v}.\nabla+\frac{1}{2}\nabla.{\bf v}\right)\psi
\label{eqn:DE1}\end{equation}
For a uniform ${\bf v}$ and a plane wave $\psi$, we have the eigenvalue equation for $\omega({\bf k})$:
\begin{equation}
(\hbar \omega({\bf k})-\hbar {\bf k}\cdot{\bf v}){\bf 1}=(B+\hbar{\bf A}\cdot{\bf k})^{2n}
\label{eqn:DE2}\end{equation}
To  determine  $B$ and $\bf A$ we assume  that the RHS matrix is a unit matrix, up to a scalar factor,
and require that 
\begin{equation} B^2=b^2{\bf 1}, A_i^2=a^2{\bf 1}, A_iA_j=-A_jA_i, i\neq j, BA_i=-A_iB, \label{eqn:DE3}\end{equation}
giving
\begin{equation}
\hbar \omega({\bf k})=(b^2+a^2\hbar^2 k^2)^n+\hbar {\bf k}\cdot{\bf v}
\label{eqn:DE4}\end{equation}
which is (\ref{eqn:PDE2}).  By the previous nLR arguments we then have $n=1/2$,  $a=c$ and $b=mc^2$, and we choose $B=mc^2\beta$, $A_i=c\alpha_i$, where $\alpha_i$ and $\beta$ are the usual Dirac matrices, to satisfy ({\ref{eqn:DE3}), so that (\ref{eqn:DE1}) becomes the generalised Dirac equation.
\begin{equation}
i\hbar\frac{\partial \psi}{\partial t}=-i\hbar\left(  c{\vec{ \alpha.}}\nabla + {\bf
v}.\nabla+\frac{1}{2}\nabla.{\bf v}  \right)\psi+\beta m c^2\psi
\label{eqn:DE5}\end{equation}
where $\psi$ is now a 4-component spinor. The above choice of the matrices $B$ and $A_i$  linearises the operator in (\ref{eqn:DE1}).  We emphasise that this derivation has been independent of the SR formalism, as it must be as SR  contains no actual relativistic effects that are not in Galilean Relativity.  So the original derivation of this equation by Dirac from SR was  fortuitous: it actually follows from neo-Lorentz Relativity. Eqn.(\ref{eqn:DE5}) automatically gives the phenomenon of gravity within the context of  the Dirac spinor. The derivation of the rest mass energy $mc^2$ has always been problematic within SR; now we understand why - it is not deriveable from SR.  Again in the limit of low $k$ (\ref{eqn:DE5}) reduces to the Pauli version of the generalised Schr\"{o}dingier equation, involving the three $2\times 2$ Pauli matrices  $\vec{\sigma}$ for spin 1/2.

Note: because of the fractally textured dynamical space, the ${\bf v}$ terms cannot be removed by a change of Frame of Reference, even locally, and which would be expected to have dynamical consequences for quantum systems, such as providing a mechanism for space-stimulated  transitions, and for extreme space fluctuations, say near a black hole, the generation of new matter, and also in the earlier moments of the Universe.

\section{Conclusions} 
It is now possible to give a definitive account of the Relativity Principle that has emerged from experiment, {\it viz} nLR, together with the dynamical space which underpins that RP.
In contrast SR is not based upon experiment, but emerges logically from GaR, but historically emerged from the mistaken belief that the speed of light in vacuum was isotropic for all observers.
Then in SR  the relativistic effects, such as length contractions, and time dilations, are purely mathematical coordinate-dependent artefacts, whereas in nLR these dynamical effects emerge from dynamics, {\it viz}, the motion of actual objects and clocks wrt the dynamical space. As discussed herein experiments have tested these SR and nLR  relativistic effects, and only nLR is in agreement with experiment.

Because clock rates, and object lengths, are affected by absolute motion, it is important to clearly specify, at least in principle, how space and time coordinates are  defined.  To that end, in nLR, 
in principle, one could measure distances and time intervals using rods and clocks, but then these measurements must be corrected for absolute motion effects, using an absolute velocity detector, with a number of practical designs being available. These corrected coordinates are  then used in the neo-Lorentz RP, which are given in (\ref{eqn:Gcoords}) and (\ref{eqn:GSpeedTransf}).  
So these definitions of coordinates depend upon the existence of  the dynamical space.  Of course these are only macroscopic coordinates; at a sufficiently small scales the concepts of space and time break down.  So the nLR RP is the same as that for GaR, but requires refinements for the absolute motion effects.  We have shown that SR is exactly GaR but using special mixed GaR space and time coordinates, constructed so that the speed of light in vacuum is isotropic, but only wrt  these special coordinates. But then there are no true relativistic effects, as no actual dynamical space is a part of SR.

Experiments, astronomical observations and theoretical discoveries have resulted in a new account of reality, with the main development being the recent discovery of space as a dynamical and fractally textured system.
This has changed our understanding of the appropriate ``Relativity Principle", which gives us the reality-determined mapping between observations by different observers, and how a Frame of Reference is, in principle, to be defined, and the ``Dynamics of Space", which now explains gravity as an emergent phenomenon and leading to an understanding of numerous  gravitational and space phenomena.  The experiments have now revealed that it is a neo-Lorentz Relativity that describes reality, together with  the recently discovered dynamical theory of space.   When setting up a FoR an observer must take account of the physics of reality, namely that the lengths of rods are affected by their absolute motion, or alternatively the distances defined by round-trip travel times depends on the observers absolute motion through space, given that we have now established that the speed of light is $c$, in vacuum, only wrt the dynamical space.. As well clocks are affected by their absolute motion. While demonstrating that GaR and SR are the same RP, it is also possible to show that the operation of the GPS is also derivable from the dynamical -3-space theory, contrary to the claim that it is a uniquely GR dependent technology.

Finally it is important to recognise the fundamental importance of the NASA spacecraft earth-flyby Doppler shift data \cite{And2008,CahillNASA}, for it has confirmed numerous laboratory light speed anisotropy experiments, in a way that is independent of the modelling of those experiments. In particular that without the Fitzgerald-Lorentz contraction the 2nd order in $v/c$ experiments would be inconsistent with the Doppler shifts, and with 1st order in $v/c$ experiments \cite{CahillFractalSpace}.  An important new result is the re-emergence of simultaneity as a meaningful actual phenomenon,  for cosmic time exist, i.e. is now measurable. The existence of cosmic time, and simultaneity, implies that the universe is significantly more interconnected than previously considered. This will have major consequences for many phenomena including cosmology. Table \ref{tab:summary} gives a comparative summary of the issues discussed.

\appendix
\section{Appendix: 
 Gas-Mode Michelson \newline Interferometer\label{append:MM}}
{\small

We derive the calibration constant $k^2$ for the Michelson-Morley interferometers in the case of Lorentzian Relativity.
For the  case of Galilean Relativity the derivation is simply repeated - without any contraction effect.  
The two arms are constructed to have the same lengths  when they are physically parallel to each other.
For convenience assume that the value $L_0$ of this length   refers to the lengths when at rest wrt space 
The Fitzgerald-Lorentz effect is that the arm $AB$  parallel to the direction of motion is shortened to
\begin{equation}
L_{\parallel}=L_0\sqrt{1-\frac{v_R^2}{c^2}}
\label{eqn:FLcontraction}\end{equation}
where $v_R$ is the speed of the arm relative to space.

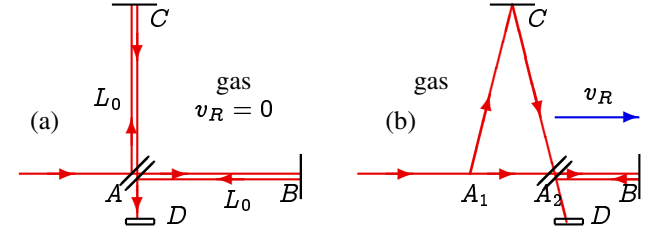
\begin{figure}[h]
\vspace{-2mm}
\setlength{\unitlength}{0.75mm}
\hspace{2mm}\begin{picture}(0,30)
\thicklines

\definecolor{hellgrau}{gray}{.8}
\definecolor{dunkelblau}{rgb}{0, 0, .9}
\definecolor{roetlich}{rgb}{1, .7, .7}
\definecolor{dunkelmagenta}{rgb}{.9, 0, .0}
\definecolor{green}{rgb}{0, 1,0.4}
\definecolor{black}{rgb}{0, 0, 0}
\definecolor{dunkelmagenta}{rgb}{.9, 0, .0}

 \color{dunkelmagenta}
  \put(10,5){\vector(0,1){5}}
\put(11,25){\vector(0,-1){5}}
\put(11,30){\line(0,-1){38}}
\put(11,-2){\vector(0,-1){5}}
\put(-10,0){\line(1,0){50}}
\put(-5,0){\vector(1,0){5}}
\put(40,-1){\line(-1,0){29.2}}
\put(15,0){\vector(1,0){5}}
\put(30,-1){\vector(-1,0){5}}
\put(10,0){\line(0,1){30}}


 \color{black}
 \put(9,-8){\line(1,0){5}}
\put(9,-9){\line(1,0){5}}
\put(14,-9){\line(0,1){1}}
\put(9,-9){\line(0,1){1}}

\put(6.5,30){\line(1,0){8}}
\put(40,-4.5){\line(0,1){8}}
 \put(40,-4.5){\line(0,1){8}}

 \put(8.0,-2){\line(1,1){5}}
\put(9.0,-2.9){\line(1,1){5}}
 \put(35,-5){ $B$}
\put(25,-6){ $L_0$}
\put(12,26){ $C$}
 \put(4,-5){ $A$}
 \put(15,-9){ $D$}

\put(2,12){ $L_0$}

 \color{dunkelmagenta}
\put(50,0){\line(1,0){50}}
 \put(55,0){\vector(1,0){5}}
\put(73,0){\vector(1,0){5}}
\put(85,0){\vector(1,0){5}}
 
\put(100,-1){\vector(-1,0){5}}

\put(68.5,-1.5){\line(1,1){4}}
\put(69.3,-2.0){\line(1,1){4}}

\put(70,0){\line(1,4){7.5}}
\put(70,0){\vector(1,4){3.5}}
\put(77.5,30){\line(1,-4){9.63}}
\put(77.5,30){\vector(1,-4){5}}

\put(83.3,-1.5){\line(1,1){4}}
\put(84.0,-2.0){\line(1,1){4}}
\put(100,-1){\line(-1,0){14.9}}

 \color{black}
\put(73.5,30){\line(1,0){8}} 
\put(100,-4.5){\line(0,1){8}}
\put(82.3,-2.5){\line(1,1){4}}
 \put(82,-2){\line(1,1){5}}
\put(83,-2.9){\line(1,1){5}}

 \put(85,-8){\line(1,0){5}}
\put(85,-9){\line(1,0){5}}
\put(90,-9){\line(0,1){1}}
\put(85,-9){\line(0,1){1}}

 \put(67,-5){ $A_1$}
\put(80,-5){ $A_2$}
\put(90,-9){ $D$}
\put(95,-5){ $B$}
\put(79,26){ $C$}
\put(89,13){ $v_R$}
\put(20,10){ $v_R=0$}
\put(-8,8){(a)}
\put(55,8){(b)}
\put(60,15){gas}
\put(25,15){gas}

\color{dunkelblau}
\put(85,10){\vector(1,0){15}}  

\end{picture}

\vspace{3mm}
\caption{\small{Schematic diagrams of the gas-mode Michelson Interferometer, with beam splitter/mirror at $A$ and
mirrors at $B$ and $C$ mounted on arms  from $A$, with the arms of equal length $L_0$ when at rest.  $D$ is the detector screen. In (a) the interferometer is at rest in space. In (b) the instrument and gas is moving through  3-space  with speed $v_R$   parallel to the $AB$ arm.  Interference fringes are observed at  $D$ when mirrors $B$ and $C$ are not exactly perpendicular -  the Hick's effect.  As the interferometer is rotated in the plane  shifts of  these fringes are seen in the case of absolute motion, but only if the apparatus operates in a gas.  By measuring fringe shifts the speed $v_R$ may be determined. } \label{fig:Minterferometer}}
\end{figure}

 \begin{figure}[t]
\vspace{0mm}
\hspace{5mm}\includegraphics[scale=0.9]{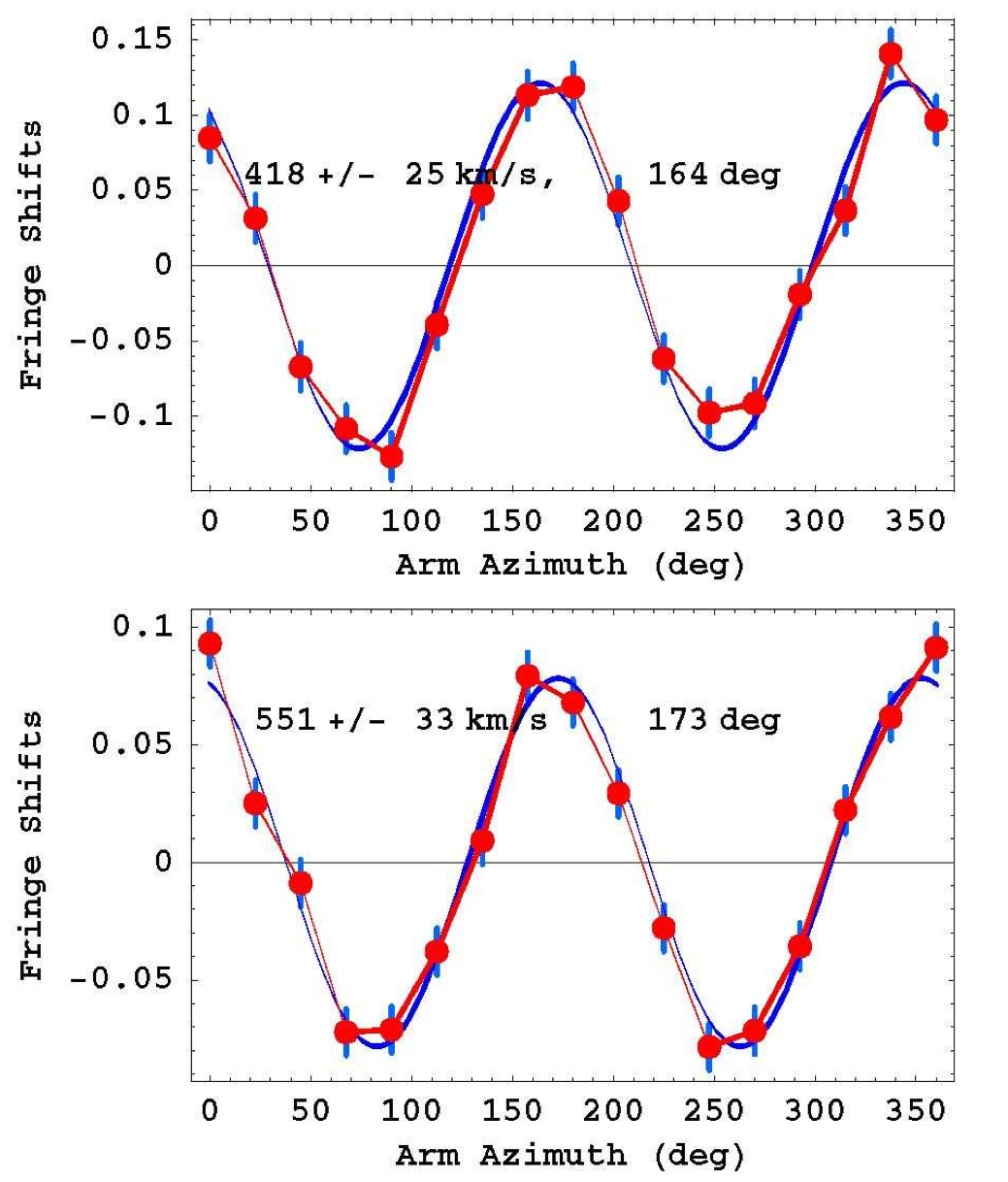}
\vspace{-2mm}\caption{\small {Top: Typical Miller data from 1925/26 gas-mode Michelson interferometer \cite{Miller}, from averaging 20 360$^\circ$ rotations.  Bottom: Data from Michelson-Morley 1887 gas-mode interferometer, from averaging 6 360$^\circ$ rotations. SR based calibration for interferometer predicts speed $<$10km/s, in disagreement with spacecraft Doppler shift data, and 1st order in $v/c$ experiments.}}
\label{fig:MandMM}\end{figure}

Following Fig.\ref{fig:Minterferometer}  let the time taken for light  to travel from
$A\rightarrow B$ be $t_{AB}$ and that from $B\rightarrow A$ be $t_{BA}$, where $V$ is the speed of light relative to the gas, which is moving with the detector. We shall also neglect the Fresnel drag effect, so $V=c/n$. 
Then
\begin{equation}
Vt_{AB}=L_{\parallel}+v_Rt_{AB} \mbox{ \ \ and  \ \ }
Vt_{BA}=L_{\parallel}-v_Rt_{BA}.
\end{equation}
\begin{eqnarray}\label{eq:ABA}
t_{ABA}=t_{AB}+t_{BA}&=&\frac{L_{\parallel}}{V-v_R}+\frac{L_{\parallel}}{V+v_R}\\
&=&\frac{2L_0V\sqrt{1-\displaystyle\frac{v_R^2}{c^2}}}{V^2-v_R^2}.
\end{eqnarray}
For the other arm, with no contraction,
\begin{equation}
\left(Vt_{AC}\right)^2=L_0^2+\left(v_Rt_{AC}\right)^2
\end{equation}
\begin{equation}
t_{AC}=\frac{L_0}{\sqrt{V^2-v_R^2}}, \mbox{\ \ \ \  }t_{ACA}=2t_{AC}=\frac{2L_0}{\sqrt{V^2-v_R^2}},
\end{equation}
giving finally for the travel time difference for the two arms
\begin{equation}
 \Delta t= \frac{2L_0V\sqrt{1-\displaystyle\frac{v_R^2}{c^2}}}{V^2-v_R^2}-\frac{2L_0}{\sqrt{V^2-v_R^2}}.
\label{eqn:MM}\end{equation}
Now trivially $\Delta t =0$  if $v_R=0$, but  also $\Delta t =0$ when $v_R\neq 0$ but only if $V=c$, {\it viz} vacuum.  This then 
  would  result in a null result on rotating the apparatus.  Hence the null result of the Michelson-Morley
apparatus is only for the special case of light travelling in vacuum. However if the apparatus is immersed in a gas then $V<c$ and a non-null
effect is expected on rotating the apparatus, since now  $\Delta t \neq 0$.  It is essential then in analysing data
to correct for this refractive index effect.  Putting
$V=c/n$ in (\ref{eqn:MM}) we find, for $v_R \ll V$ and  when $n \approx 1$, that 
\begin{equation}
\Delta t= n(n^2-1)\frac{L_0v_R^2}{c^3}.
\label{eqn:LCalib}\end{equation}
 However if the data is analysed not using the
Fitzgerald-Lorentz contraction (\ref{eqn:FLcontraction}), then, as done in the old analyses,   the estimated time difference is 
\begin{equation}
\Delta t = \frac{2L_0V}{V^2-v_R^2}-\frac{2L_0}{\sqrt{V^2-v_R^2}},
\end{equation}
which again for $v_R \ll V$  gives 
\begin{equation}
\Delta t =n^3\frac{L_0v_R^2}{c^3}.
\end{equation}
With Fresnel drag and $n\approx 1$, the sign of $\Delta t$ in (\ref{eqn:LCalib}) is reversed \cite{Book}.
Symmetry arguments easily show that when rotated we obtain a $\cos(2\theta)$ factor in (\ref{eqn:LCalib}).  When analysing the data the temperature induced drift in $L$, and the Hick's effect, need to be taken into account, see \cite{CahillNASA}.  Examples of fringe shift data from Michelson and Morley, and Miller, are shown in Fig.\ref{fig:MandMM}, with speeds determined  using  (\ref{eqn:LCalib}). These speeds agree with speeds from 1st order experiments, and from the spacecraft earth-flyby Doppler shift data.
}

\section{Appendix:  Clock Slowing in  \newline neo-Lorentz Relativity\label{append:propertime}}
The two arms of the Michelson interferometer act as two independent ``photon" clocks, with one ``click" defined to be the round-trip travel time.  In vacuum $V=c$, then from above,
\begin{equation}
\Delta t = \frac{2L_0}{\sqrt{c^2-v_R^2}}= \frac{\Delta t_0}{\sqrt{1-v_R^2/c^2}},
\end{equation}
for each arm, 
where $\Delta t_0=2L_0/c$ is the travel time when $v_R=0$.  So  clocks moving wrt space are slowed. Then moving-clock time, as measured by ``clicks",  is slower for a moving clock, and we can define the elapsed or proper time of such a moving clock to be 
\begin{equation}
\Delta\tau=\Delta t\sqrt{1-v_R^2/c^2}.
\label{eqn:propertimeb}\end{equation}
As shown in Sect.\ref{sect:dynamics} when $v_R$ is time dependent and/or inhomogeneous (\ref{eqn:propertime}) results in refraction of quantum matter waves, and also EM waves.  This is  the explanation for gravity.

However this clock slowing effect is not universal, for we can construct a ``photon" clock with a gas present.  Then  from the above the round-trip travel time now depends on the angle between the clock arm and the direction of motion of the clock through space.  It is important to understand that it is clocks that are affected by motion through space. Intrinsic or cosmic time may be determined by observing clock time, as defined say by (\ref{eqn:propertimeb}), and then correcting for the  absolute motion effect by independently measuring $v_R$, using say a 1st order in $v_R/c$ detector.

\section{Appendix:  Different Fresnel Drag in SR and LR}
{\small The Fresnel drag  effect is another case where LR and SR differ, and where experiment agrees only with the LR prediction.  The Fresnel drag
is a phenomenological observation that gives the speed  $V_D$, measured wrt  a dielectric (with refractive index $n$),   when the dielectric has speed $v_R$ wrt space:
\begin{equation}
V_D=\frac{c}{n}+v_R\left(1-\frac{1}{n^2}\right)
\label{eqn:Fresnel}\end{equation}
This is  confirmed  by the Optical-Fibre - RF Coaxial-Cable experiment \cite{CahillFractalSpace}, by Ring-Laser experiments that detect the sidereal effect, see \cite{CahillNASA}, and by the operation of Optical-Fiber Gyroscopes. 
However there is a spurious ``derivation" of this using SR.  Using the SR choice of coordinates in (\ref{eqn:mixedST}) leads to the ``composition law of velocities" in (\ref{eqn:SRVelTran}), giving to 1st order in $W'\!/c$ 

\begin{equation}
W=W'+\overline{V}\left(1-\frac{W'^2}{c^2}\right)
\label{eqn:1stSpeedCompa}\end{equation}
where $W'$ is the speed of an entity  wrt $O'$, and $W$ is the speed wrt $O$.
If we apply this to light in a dielectric at rest wrt $O'$,  then the light has speed $W'=c/n$ wrt observer $O'$, according to SR, which implies that 
 the speed of light wrt $O$ is then
\begin{equation}
W=\frac{c}{n}+\overline{V}\left(1-\frac{1}{n^2}\right),
\label{eqn:1stSpeedCompb}\end{equation}
which has the form of (\ref{eqn:Fresnel}), but actually has a completely different meaning,
for here $\overline{V}$ is the speed of the dielectric wrt $O$, whereas in (\ref{eqn:Fresnel})
$v_R$ is the speed of the dielectric wrt space. Then if the dielectric is at rest wrt $O$, $\overline{V}=0$, and  (\ref{eqn:1stSpeedCompb}) predicts no Fresnel drag effect, whereas 
in  Optical-Fibre Gyrocompasses and the  1st order in $v/c$ Optical-Fibre - RF Coaxial-Cable Detector, where the Fresnel drag plays a key role, the observer is at rest wrt the dielectric.  Most importantly that (\ref{eqn:1stSpeedCompb})
is wrong  is that the Fresnel drag effect is not present in RF coaxial cables \cite{CahillFractalSpace}, whereas (\ref{eqn:1stSpeedCompb}) makes no such distinction.   So SR fails to  give the observed properties of the Fresnel drag effect.  This is to be expected, since in SR it is purely a coordinate effect, and not an actual dynamical effect. 

\section{Appendix: Twin Effect}
The twin effect is that a clock $C_2$ making a round trip journey, ${\bf r}_0(t)$,   will, when returning to the ``stay-at-home clock" clock $C_1$, be retarded wrt $C_1$. However this is not generally true within nLR, as this description of $C_1$ is not well defined within nLR, because both $C_1$ and $C_2$ could be in motion wrt space, and the space may not have a uniform velocity wrt either observer.   
The elapsed proper time for $C_2$ is
\begin{equation}\Delta \tau_2=\int_0^T dt \sqrt{1-\left(\frac{d {\bf r}_o}{dt}-\bf{v}({\bf r}_0(t), t)\right)^2/c^2} 
\end{equation}
where ${\bf v}_R={d {\bf r}_o}/{dt}-\bf{v}({\bf r}_0(t), t)$  is the velocity of $C_2$ wrt the local space, and position and time are cosmic coordinates, and the time $T$ is defined by  $\int_0^T dt\,{\bf r}_0(t)={\bf 0}$ - the round-trip condition.  A similar expression holds for $C_1$.  Special circumstances are needed to obtain the ``twin effect":   Suppose that the space velocity at $C_1$ is zero, which maximises the $C_1$ elapsed time,
then 
\begin{equation}\Delta \tau_1=\int_0^T dt
\end{equation}
and it trivially follows that $\Delta \tau_2 < \Delta \tau_1=T$, which is the twin effect.  This effect is a consequence of absolute motion wrt space.

}

\newpage
\footnotesize{

}


\begin{thebibliography}{99}

\bibitem{MMCK} Cahill R.T. and Kitto K. {\it Michelson-Morley Experiments Revisited}, {\it Apeiron},  2003, v. 10(2),104-117.
\bibitem{MMC}  Cahill  R.T. {\it The Michelson and Morley 1887 Experiment and the Discovery of Absolute Motion},   {\it Progress in Physics}, 2005,  v. 3, 25-29.

\bibitem{Mink}  Minkowski H. ``Raum und Zeit'', {\it Physikalische Zeitschrift}, 10  {\it Jahrgang},  104-115, 1909.  English translation in {\it The Principle of Relativity} by Lorentz H.A., Einstein A., Minkowski H. and Weyl H., translated by Perrett W. and Jeffery G.B., Dover, New York, 1952.

\bibitem{Book} Cahill  R.T. { \it Process Physics: From Information Theory to Quantum Space and Matter},  Nova Science Pub., New York, 2005.      
\bibitem{Review}  Cahill R.T. {\it Dynamical 3-Space: A Review},   in {\it Ether Space-time and Cosmology: New Insights into a Key Physical Medium},   Duffy M. and L\'{e}vy  J., eds.,  {\it Apeiron}, 135-200, 2009.  

\bibitem{Paradigm}  Cahill R.T. {\it Unravelling the Dark Matter - Dark Energy Paradigm}, {\it Apeiron},  2009, v. 16(3),  323-375.
\bibitem{EmergentGravity}  Cahill R.T. {\it Dynamical 3-Space: Emergent Gravity}, in {\it Should the Laws of Gravitation be Reconsidered?},  M\'{u}nera H.A. ed. (Montreal: Apeiron 2011).
 
\bibitem{BlackHoles}  Cahill R.T. and  Kerrigan D.  {\it  Dynamical Space: Supermassive Galactic Black Holes and Cosmic Filaments},   {\it Progress in Physics}, 2011,  v. 4, 65-68.

\bibitem{Universe}  Cahill R.T. and  Rothall D.  {\it  Discovery of Uniformly Expanding Universe},   {\it Progress in Physics},  2012, v. 1,  65-68.
     
\bibitem{CahillNASA}  Cahill R.T.    {\it Combining NASA/JPL One-Way Optical-Fiber Light-Speed Data with Spacecraft Earth-Flyby Doppler-Shift Data to Characterise 3-Space Flow},   {\it Progress in Physics}, 2009, v. 4,  50-64.

\bibitem{CahillFractalSpace}  Cahill R.T.    {\it Characterisation  of Low Frequency Gravitational Waves from    Dual RF Coaxial-Cable Detector: Fractal Textured Dynamical 3-Space},   {\it Progress in Physics},  2012, v. 3, 3-10.
 
\bibitem{And2008} Anderson J.D., Campbell J.K., Ekelund J.E., Ellis J. and Jordan J.F. {\it  Anomalous Orbital-Energy Changes Observed during Spacecraft Flybys of Earth},  {\it Phys. Rev. Lett.},  2008, v. 100,  091102.



\bibitem{BrownContraction}  Brown H.R. {\it The Origins of Length Contraction; I The Fitzgerald-Lorentz Deformation Hypothesis},   {\it Am. J. Phys.}, 2001,  v. 69, 1044-1054.

\bibitem{BrownBook}  Brown H.R. {\it Physical Relativity: Space-Time Structure from a Dynamical Perspective},   {\it Clarendon Press Oxford}, 2005.

\bibitem{BrownPooley}  Brown H.R. and Pooley O. {\it Minkowski Space-Time: a Glorious Non-Entity},  in {\it The Ontology of Spacetime}, Dieks, ed. {\it Elsevier} 2006, 67-89.

\bibitem{Hertz}  Hertz, H.  {\it On the Fundamental Equations of Electro-Magnetics for Bodies in Motion},  {Wiedemann's Ann.}, 1890,  v. 41, 369;  {\it  Electric Waves, Collection of Scientific Papers},   {\it Dover Pub., New  York},  1962.

\bibitem{Lorentz}  Lorentz H.A.  {\it 	Electromagnetic Phenomena in System Moving with any Velocity Smaller than that of Light }, 
	Proc. of the Royal Netherlands Academy of Arts and Sciences, 1904, v. 6,  809-831.
\bibitem{Lorentz1892}  Lorentz H.A. {\it De relatieve beweging van de aarde en den aether}, Amsterdam, Zittingsverlag Akad. v. Wet., 1892, 1,   p. 74-79, (transl.: The relative motion of the earth and the aether).
 \bibitem{Fitzgerald} Fitzgerald G.F.  {\it  The Ether and the Earth's Atmosphere}, {\it Science},  1889, v. 13,   390.
\bibitem{CahillMink}  Cahill R.T.    {\it Unravelling Lorentz Covariance and the Spacetime Formalism},   {\it Progress in Physics},  2008, v. 4,   19-24.

\bibitem{Miller}   Miller D.C. {\it  The Ether-Drift Experiment and the Determination of the Absolute Motion of the Earth}, {\it Rev. Mod. Phys.},  1933, v. 5, 203-242.

\bibitem{CahillSE}  Cahill R.T.    {\it  Dynamical Fractal 3-Space and the Generalised Schr\"{o}dinger Equation: Equivalence Principle and Vorticity Effects},   {\it Progress in Physics},  2006, v. 1, 27-34.


\end{thebibliography}
\end{document}